\documentclass[9pt,shortpaper,twoside,web]{ieeecolor}
\usepackage{generic}
\usepackage{cite}
\usepackage{amsmath,amssymb,amsfonts}
\usepackage{algorithm}
\usepackage{algpseudocode}
\usepackage{graphicx}
\usepackage{multirow}
\usepackage{siunitx}
\usepackage{textcomp}
\def\BibTeX{{\rm B\kern-.05em{\sc i\kern-.025em b}\kern-.08em
    T\kern-.1667em\lower.7ex\hbox{E}\kern-.125emX}}
\begin{document}

\title{Medical Profile Model: Scientific and Practical Applications in Healthcare}
\author{Pavel Blinov and Vladimir Kokh
\thanks{Pavel Blinov and Vladimir Kokh are with the Sber Artificial Intelligence Laboratory, Moscow, Russia (e-mail: blinoff.pavel@gmail.com; kokh.v.n@gmail.com).}}

\maketitle

\begin{abstract}
The paper researches the problem of representation learning for electronic health records. We present the patient histories as temporal sequences of diseases for which embeddings are learned in an unsupervised setup with a transformer-based neural network model. Additionally the embedding space includes demographic parameters which allow the creation of generalized patient profiles and successful transfer of medical knowledge to other domains. The training of such a medical profile model has been performed on a dataset of more than one million patients. Detailed model analysis and its comparison with the state-of-the-art method show its clear advantage in the diagnosis prediction task. Further, we show two applications based on the developed profile model. First, a novel Harbinger Disease Discovery method allowing to reveal disease associated hypotheses and potentially are beneficial in the design of epidemiological studies. Second, the patient embeddings extracted from the profile model applied to the insurance scoring task allow significant improvement in the performance metrics.
\end{abstract}


\section{Introduction} \label{introduction_sect}
Nowadays, Electronic Health Records (EHRs) are an essential part of each modern national healthcare system, as it provides quick access to patient-related information for all interested parties (the patient itself, an attending physician and medical staff, regulatory authorities, etc.). While a person can manually analyze a single EHR data file, this task becomes tedious and labor-intensive as the number of records scales up. This problem becomes especially obvious on a national, regional, or even individual clinic scale. The prioritization of patients for a health check-up or prediction of disease complications are good examples of such problems. Therefore automatic intellectual tools are required to facilitate effective interaction and comprehensive analysis of EHRs~\cite{dash2019big}.

Even more challenging problems emerge beyond prognostic models and simple management operations with health records. More specifically, what knowledge (and in what form?) can be extracted from a database of EHRs, and how (if possible?) it can be transferred to another domain? This problem is actively researched in the deep domain adaptation area as current neural networks have a large capacity to learn transferable and valuable representations between the source and target domains~\cite{xu2020transfer}.

In this paper, we address both of the specified problems. We treat EHR data in the current paradigm from the Natural Language Processing (NLP) area and develop the model for encoding diagnosis events (defined with International Classification of Diseases codes, ICD-10~\cite{icd10}) as embedding vectors. With such a model build, we can move all patients to the latent embedding space where the above analytical and forecasting tasks can be easily solved.

Moreover, due to our way of embedding construction, they acquire alienable and reusable properties, which closes the second group of posed questions about the possibility of knowledge transferring between domains. To practically demonstrate this, we select the health insurance domain as the target while keeping medical as the source. The main issue is that datasets from both domains are fully anonymized and unrelated, containing no linkage keys between each other. However, even in such a setup, we propose the solution by extracting medical patient representations and exploiting them in the target domain.

To summarize, the main contributions of this paper are the following:
\begin{itemize}
    \item We develop the medical profile model based on the neural network transformer architecture simultaneously handling demographic and temporal data about diagnosis events. The model differs from the analogies by the inclusion of learnable gender embeddings. This feature of the model allows inferring generalized patient representations useful in the downstream tasks.
    \item We perform a comprehensive analysis of the profile model and resulting embeddings. Its performance is compared with several internal baselines and the state-of-the-art model.
    \item We propose a novel data-driven \emph{Harbinger Disease Discovery} method for disease-associated hypotheses search, which potentially leads to more consistent and intelligent exploratory medical research. The candidate hypotheses selection is based on conditional risk probability comparison for control and simulated patient groups. The method utilizes the profile model as a proxy for the medical risk estimate.
    \item We show a successful case of knowledge transferring between ill-related domains. The generalized patient embeddings from the profile model significantly improve the performance in the task of insurance risk scoring (ROC AUC metric improved by 1.82\%).
\end{itemize}

The rest of this paper is structured as follows.
Section~\ref{related_work_sect} provides an overview of related works for embedding learning in the medical domain. The description of the used datasets is given in Section~\ref{Data}. The model development process and experimental results are discussed in Section~\ref{profile_model_sect}. Section~\ref{applications_sect} is devoted to the applications of the profile model. In the last Section~\ref{Conclusions}, we provide concluding remarks and list directions for future work.

\begin{table*}
\centering
\caption{Comparison of profile model with analogies.} \label{model_comparison}
%\begin{tabular}{lllll}
\begin{tabular}{m{2.2cm} m{3.2cm} m{3.2cm} m{3.2cm} m{3.2cm}}
\hline
\bfseries Criteria & \bfseries Profile model & \bfseries BEHRT & \bfseries Med-BERT & \bfseries MusaNet \\
Pretraining data source & DMed (general EHR) & CPRD (primary care data)~\cite{herrett2015data} & Cerner Health Facts (general EHR) & MIMIC III (ICU data)~\cite{johnson2016mimic} \\
\hline
Pretraining task & Masked LM & Masked LM & Masked LM + prediction of prolonged length of stay in hospital & Not used \\
\hline
Total number of pretraining patients & 1M & 1.6M & 20M & 7.5K \\
\hline
Type of input code & ICD-10 & ICD-10 & ICD-9 + ICD-10 & ICD-9 \\
\hline
Vocabulary size & 7,096 & 301 & 82K & 4,874 \\
\hline
Input structure & Code + visit + age embeddings + gender embeddings & Code + visit + age embeddings & Code + visit + code serialization
embeddings & Code + visit
embeddings \\
\hline
Evaluation tasks & Next diagnosis prediction, Medical risk scoring, Insurance scoring & Multi-disease prediction & Disease predictions according to
strict inclusion/exclusion criteria & Next diagnosis prediction, Future re-admissions \\
\hline
\end{tabular}
\end{table*}

\section{Related Work} \label{related_work_sect}
To a great extent, the substantial part of the machine learning area is about finding "good" representations for modeling objects. The current development of this problem leads to the automatic methods of feature engineering with neural networks (opposite to hand-crafted methods). Such feature vectors for modeling objects are called embeddings~\cite{cai2018comprehensive}. They are mainly trained in an unsupervised fashion to capture intrinsic relations between objects of a target domain, for example, word ordering and associations in natural language text.

In the medical field, the above-mentioned task is often treated as embedding learning for the domain concepts~\cite{bai2019medical,ChoiCS16}, implying ICD diagnosis codes, terms, abbreviations, medication and procedure names, etc. Commonly such concepts are viewed within a temporal treatment process associated with a patient. For example, a patient has a history of clinical visits; each visit has several tagged diseases and procedures, prescribed medications, etc. In the current paper, we follow a similar modeling approach. But in favor of our practical tasks, we intentionally restrict the type of learnable concepts only to visits, diagnosis codes, and the most basic demographic attributes of a patient (gender and age). Although this reduces the expressive power of our model but makes it language-independent. Besides, such a model implies more anonymity as no textual data have involved and no confidential information can be leaked which eases the model deployment process.

To efficiently deal with sequentially organized medical data the notion of context becomes crucial. For this, some papers~\cite{bai2019medical,ChoiCS16} use modifications of the Skip-gram algorithm~\cite{mikolov2013} though it is limited to account for only a fixed-size context of a sequence. Models with Recurrent Neural Network (RNN) architectures offer a better context-handling mechanism. There are whole generations of RNN-based models for a patient representation task~\cite{doctorAI,dipole,deepCare}. We also benchmark this type of model but only as a baseline system. The bottleneck of RNNs is the single internal state vector that has to retain all information about the sequence. More advanced and powerful architectures (like current attention-based models~\cite{vaswani2017attention}) can process the whole sequence context more elaborately. The model with transformer-based architecture is our central focus in this work. There are quite a few studies related to the application of transformer-like models to ICD code predictions based on a variety of clinical text types (medical notes, case studies, discharge summaries, etc.)~\cite{blinov_aime2020,DL4ICD_2020comparison,XLM4ICDClassification}. In this paper, we try to abstract from atomized visits and model the entire patients' histories to research what patterns can be learned from such patient tracks. In such formalization, our proposed model is most similar to BEHRT~\cite{li2020behrt}, Med-BERT~\cite{Med_BERT}, and MusaNet~\cite{Peng20}. We try to summarize the comparison between our and these models in Table~\ref{model_comparison}.

Firstly, all similar models use the Bidirectional Encoder Representations from Transformers (BERT) architecture~\cite{devlin2019bert}. All models, except the MusaNet, resort to the pre-training technique with Masked Language Modeling (MLM) followed by fine-tuning for the downstream tasks. As can be noted from the last row of Table~\ref{model_comparison}, our work is favorably distinguished by the variety of practical applications. Furthermore, in the task of next diagnosis prediction we do not restrict to the specific diseases as in Med-BERT with a prediction of heart failure and the onset of pancreatic cancer. In terms of the data source, the profile model is closest to the Med-BERT but inferior to it regarding data size.

Standardization and structuring of such a complex domain as medicine is not an easy task. Unfortunately, this leads to weak compatibility between ICD versions. The transition process from the previous ICD-9 standard is still ongoing in many countries, even though the updated ICD-10 version has endorsed by the World Health Organization (WHO) in 1990! That means that many research projects and papers still proceeded in terms of obsolete ICD-9 standards. Unlike Med-BERT and MusaNet we mainly experiment with modern ICD-10 resorting to the 9th version only for model comparison purposes (see Subsection~\ref{mimic_sect}).

\begin{table}
\centering
\caption{Datasets statistics.} \label{table_data}
\begin{tabular}{rccc}
\hline
\multirow{2}{*}{\bfseries Decade}
& \multicolumn{3}{c}{\bfseries Count (Male, \%)}\\
\cline{2-4}
& \bfseries DMed & \bfseries MIMIC-III & \bfseries DIns \\
\hline
{[0; 10)} & 142,549 (51.8) & 228 (52.2) & 19,212 (48.3)\\
{[10; 20)} & 119,176 (51.3) & 10 (80.0) & 45,332 (51.3)\\
{[20; 30)} & 183,836 (49.1) & 181 (50.8) & 5,522,892 (46.9)\\
{[30; 40)} & 165,383 (50.3) & 325 (54.2) & 9,497,249 (52.0)\\
{[40; 50)} & 127,488 (50.0) & 818 (57.9) & 8,215,431 (59.4)\\
{[50; 60)} & 146,516 (46.2) & 1,276 (59.2) & 8,577,933 (65.2)\\
{[60; 70)} & 101,071 (41.9) & 1,613 (58.5) & 4,603,950 (63.2)\\
{[70; 80)} & 50,555 (33.2) & 1,652 (55.6) & 322,176 (54.7)\\
{[80; 90)} & 24,265 (24.1) & 1,393 (48.4) & 31,170 (50.2)\\
{[90; 100)} & 2,650 (16.0) & 0 (n/a) & 0 (n/a)\\
\hline
Total & 1,063,489 & 7,496 & 36,835,345\\
\hline
\end{tabular}
\end{table}

Among the models from Table~\ref{model_comparison}, the profile model is the second by vocabulary size of learnable concepts. Among the listed models only our one introduces learnable embeddings for a gender. In conjunction with age embeddings, this leads to a significant boost in model performance on the task of the next diagnosis prediction prediction (see Section~\ref{mimic_sect}). Moreover, such a feature allows the extraction of valuable patient representations useful in downstream practical applications (see Section~\ref{ins_scoring}).

\section{Data} \label{Data}
At our disposal, we had three anonymized datasets: MIMIC-III~\cite{johnson2016mimic}, our private medical (\emph{DMed}) and insurance (\emph{DIns}) data\footnote{The use of \emph{DMed} and \emph{DIns} datasets has been approved by the Independent Ethics Committee of the Moscow Regional Branch of the Russian Society of Roentgenologists and Radiologists (Protocol No. 04/2022 dated April 21, 2022).}. We treat \emph{DMed} as the primary data for model development and use \emph{DIns} to estimate the performance of embeddings in the scoring task (see Subsection~\ref{ins_scoring}). Compared to our in-house datasets, MIMIC-III is rather a toy data, but it is the only open-source alternative to compare our model with the others.

In both domains, for modeling purposes, we operate on a person (patient or customer) level of abstraction. Table~\ref{table_data} lists summary statistics for three datasets by age decades. We pre-process the medical data closely following~\cite{Peng20}: removing rare ICD diagnosis codes by the threshold of 5 and leave only patients with at least two visits. For \emph{DMed} and MIMIC-III data, the average numbers of visits are 9.9 and 2.66, respectively.

The \emph{DIns} dataset counts 36,835,345 applicants for 8 years (from 2013 to 2020). The \emph{DMed} data comes from a net of region-level clinics, in total including 1,063,489 patients for the 6 years (from 2014 to 2019). For the \emph{DMed} data, we left a random 5\% sample of 53,175 patients for validation and analysis. Insurance data was split by time into three parts: \emph{train} (2013-2017), \emph{validation} (2018), and \emph{test} (2019 and 2020). In the insurance model development, we used \emph{train} and \emph{validation}, leaving the \emph{test} for final performance estimation.

\section{Medical Profile Model} \label{profile_model_sect}
In this work, we model a patient's medical history through contextualized embeddings built with transformer architecture neural network model~\cite{devlin2019bert}. By analogy from natural language tasks such model process sequentially organized data samples. Only a sample, instead of being a text consisting of tokens (words), in our case represents a patient history consisting of medical events in the form of ICD-10 codes.

\begin{figure}[!t]
\centerline{\includegraphics[width=\columnwidth]{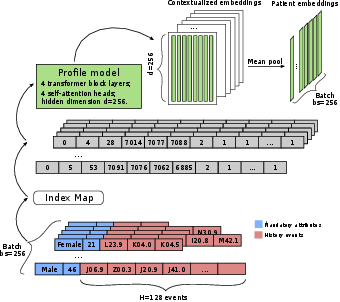}}
\caption{Scheme of data flow.} \label{figure_arch}
\end{figure}

Demographic factors as well significantly affect a person's health condition and should be included in the modeling process. In our setup, we design special tokens for both genders and each age in years from 0 to 99 (see Fig.~\ref{figure_arch} for the complete sample example). In our medical dataset, we estimate the age of a patient at the moment of their last diagnosis event. In the insurance data, we know a person's age at the moment of an insurance application.

The history lengths can significantly vary across the patients. We limit the maximum length with $H=128$ event slots and do not remove repeating codes. By this number, we can encompass more than 99\% of histories without trimming. We try to work with only well-represented ICDs and filter out codes that encounter less than 5 times across our medical data. This procedure left us with 6,986 ICD-10 codes. Thus the total length of our "token" vocabulary $V$ (with gender/age and auxiliary tokens) is $|V|=7,096$.

Figure~\ref{figure_arch} outlines the data flow pipeline in our model. At the image bottom (in blue and red colors), a couple of history samples are presented. As the raw input samples are of different lengths, on the next step they are padded to the maximum length and converted to indices (central gray-colored part) with the \emph{Index Map}.

As mentioned in Section~\ref{Data}, we use for the training 95\% of patient histories (1,010,314), counting 15,522,698 ICD events. Compared to natural language data this is a rather small dataset, so we set up the following model parameters: the dimension of embedding space $d=256$, 4 encoder layers, and 4 self-attention heads (i.e. 3 times smaller than language BERT base model~\cite{devlin2019bert}). For the decoder part, the linear projection layer ($d\times|V|$) was used. Such a model was implemented with the PyTorch framework~\cite{paszke2019pytorch} and Transformers library~\cite{Wolf2019}. We trained this model on the masked token prediction task (with 25\% mask probability parameter) with AdamW optimizer (learning rate of \num{5e-5}) and batch size ($bs=256$) of samples for 30 epochs.

At the end of training the result 256-dimensional embedding space contains vectors for each ICD code as well as genders and ages ones. And a list of contextualized vectors ${h\textsubscript{1},...,h\textsubscript{k}}$ can be produced for an input sample of $k$-length history. To go from ${h\textsubscript{1},...,h\textsubscript{k}}$ vectors to a patient embedding ${h\textsubscript{p}}$ we resort to mean pooling operation along each of $d=256$ dimensions over ${1,...,k}$ vectors. This is schematically depicted at the top part of Fig.~\ref{figure_arch} (green).

\subsection{Performance Baselines} \label{emb_analysis}
To assess the quality of the trained model, we performed several experiments.

First, on the medical validation set (see Section~\ref{Data}), we estimate the model's ability to predict the next patient's ICD code (one of 6,986) from a prefix of mandatory parameters and previous code history. We can easily do this by using the same decoder layer used for training as it returns the normalize probability distribution over all $V$ vocabulary elements from which the most probable code can be selected. By comparing actual and predicted values we compute the \emph{Accuracy} metric~\cite{Manning} for history length thresholds $th=2..64$. There are too few validation samples after $th=64$ to report accuracy. For example, for $th=12$ we trim all histories at this length and drop shorter samples. Codes at the 12th position become our ground truth values and, for left samples, we predict the most probable codes from their 11 previous events.

\begin{figure}[!t]
\centerline{\includegraphics[width=\columnwidth]{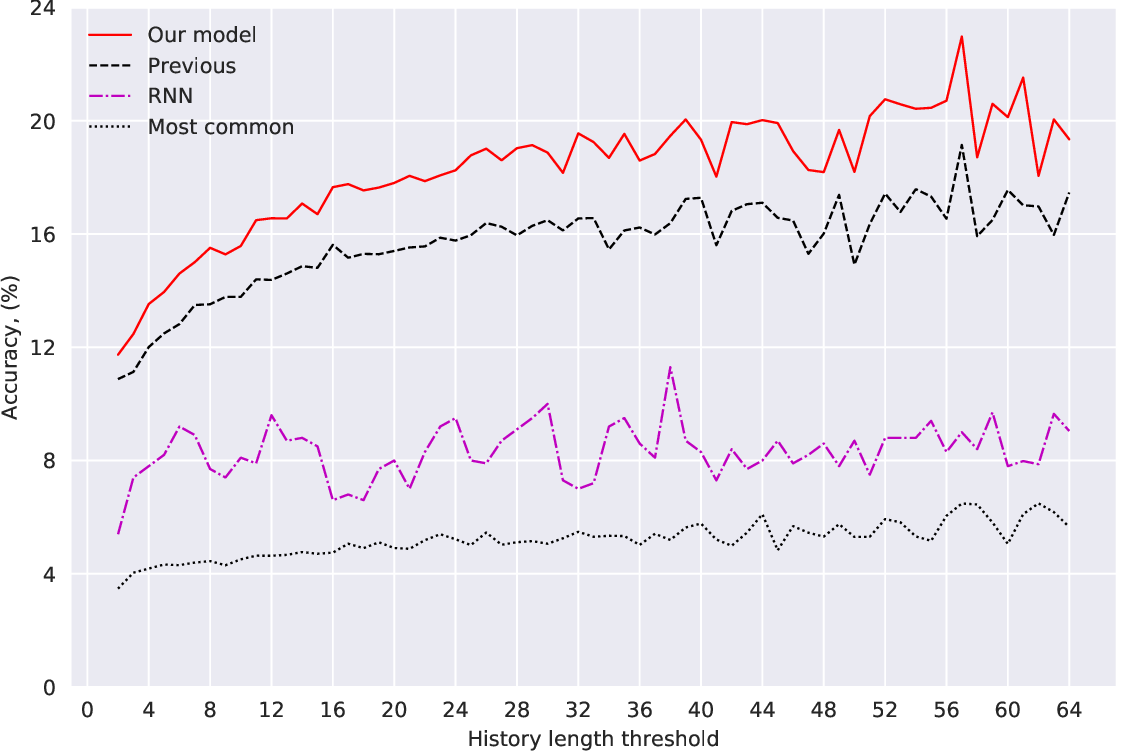}}
\caption{Model and baselines performance in the next ICD code prediction task, (\%).} \label{figure_perf}
\end{figure}

Also for comparison purposes, we implemented three baseline algorithms: \emph{Most common}, \emph{Previous} and \emph{RNN}. The \emph{Most common} baseline always predicts the constant code of most popular disease in the train data – \emph{J06.9 - Acute upper respiratory infection, unspecified}. The \emph{Previous} baseline repeats the last seen value for a sample. For the \emph{RNN} baseline we train the neural network from~\cite{doctorAI} for 30 epochs with an embedding size of 256 and 512 dimensions of gated recurrent units in two hidden layers. The performances of all algorithms are shown on Fig.~\ref{figure_perf}. It is interesting to note that the \emph{Previous} baseline shows surprisingly high performance because we keep repeating codes in histories.

\subsubsection{MIMIC-III benchmark} \label{mimic_sect}
Moreover, we compare our model to available state-of-the-art methods. Many medical models~\cite{Peng20,qiao2019mnn,song2018attend} benchmark their performance on the open source MIMIC-III \cite{johnson2016mimic} dataset in the typical task of next diagnosis prediction. This dataset uses ICD-9 diagnosis coding scheme. As mentioned in Section~\ref{related_work_sect} there is weak compatibility between ICD-10 and ICD-9 standards. So we cannot directly apply our model and had to retrain it with these data and slight modifications in the architecture\footnote{This code is available at \underline{https://github.com/pavel-blinov/mimic.profile}}.

Predictions in the MIMIC-III benchmark often have to be made not for disease codes themselves but categories in their hierarchical grouping~\cite{doctorAI,Peng20}, i.e. some form of medical risks. Following this practice, we use second-level categories from the Clinical Classifications Software\footnote{\underline{https://www.hcup-us.ahrq.gov/toolssoftware/ccs/ccs.jsp}} as possible model outcomes. This procedure reduces output space from thousands of codes to 136 categories.

A common metric to evaluate models in such a setting is the \emph{Precision at k}, which measures the percentage of relevant categories in the \emph{k} retrieved candidates:
\begin{equation}
Precision@k = \frac{|top_{k} \cap \hat{y}|}{min(k,|\hat{y}|)}, \label{precision_eq}
\end{equation}
where $top_{k}$ is the set of predicted categories, $\hat{y}$ is the set of actual categories in the next patient visit.

Given the small data size instead of a single validation split, we perform a 10-fold cross-validation procedure~\cite{bishop2006pattern} to estimate the mean and standard deviation of $Precision@k$ for $k=[5, 10, 20]$ values. Table~\ref{table_diag} shows the results for several variants of our model. Also as a reference, we include the result of \emph{MusaNet} from~\cite{Peng20} and \emph{BiteNet} from~\cite{bitenet}, which claims to be state-of-the-art in this dataset and task.

\begin{table}
\centering
\caption{Systems performances in the next diagnoses prediction task on MIMIC-III data, (\%).} \label{table_diag}
\begin{tabular}{rccc}
\hline
\multirow{2}{*}{\bfseries System}
& \multicolumn{3}{c}{\bfseries Precision@k} \\ \cline{2-4}
& \bfseries k=5 & \bfseries k=10 & \bfseries k=20 \\
\hline
MusaNet~\cite{Peng20} & 65.07 & 60.69 & 71.04 \\
BiteNet~\cite{bitenet} & 66.15$\pm$0.24 & 60.19$\pm$0.56 & 71.04$\pm$0.31 \\
v.CLS & 61.92$\pm$1.11 & 57.3$\pm$1.03 & 67.77$\pm$0.79 \\
v.cmm\_wo\_gender/age & 65.86$\pm$0.94 & 60.98$\pm$0.6 & 71.7$\pm$0.36 \\
v.cmm & 66.41$\pm$0.99 & 61.62$\pm$0.87 & 72.07$\pm$0.64 \\
v.cmm\_wo\_positional & \bf66.77$\pm$0.84 & \bf61.9$\pm$0.99 & \bf72.58$\pm$0.83 \\
\hline
\end{tabular}
\end{table}

The default choice in transformer-like models is the use of special classification (CLS) token representation~\cite{devlin2019bert} as the concise summary representation of a whole sample before final classification head. But in this task such version of the model (\emph{v.CLS}) is actually performing poorly. The \emph{MusaNet}'s result is overcome by using more advanced pooling strategies over contextualized embeddings proposed in~\cite{blinov_aime2020}. Such a model (\emph{v.cmm\_wo\_gender/age}) achieves comparable results (given the standard deviation) even without gender and age embeddings. But it is still inferior or marginally better than \emph{BiteNet}. Only inclusion of gender and age embeddings (\emph{v.cmm}) allows to outperform the \emph{BiteNet}'s result. This proves the crucial importance and novelty of these features in proposed model on the given task of next diagnosis estimation.

We also found that the transformer model without the positional embedding layer (\emph{v.cmm\_wo\_positional}) performs even better. From this, we can conclude that in the risk prediction task, the composition of patient diseases is more important than their specific ordering. Overall, the above ablation experiments confirm that the proposed model allows to achieve state-of-the-art performance in the task of subsequent patient risk prediction.

\subsection{Embeddings analysis} \label{emb_analysis}
Beside predictive power we want our embeddings to be interpretable, i.e. that our model will learn some meaningful things, like appropriate relations between ICD codes. For example, by retrieving most similar (with cosine similarity measure~\cite{Manning}) embeddings to the \emph{J06.0}-embedding we obtain close and related diseases: \emph{J04.2}, \emph{J04.0}, \emph{J20.8}, \emph{J03.8}, or \emph{J20.0}. As our age concepts are presented in the same embedding space we can ask ICD-age related questions, like what ages closest to the \emph{M41.1} code? And the nearest age values would be 14, 15, 16, 12, or 11, which seems correct as the \emph{M41.1} is \emph{Juvenile and adolescent idiopathic scoliosis}, meaning it is the adolescent disease.

\begin{figure}[!t]
\centerline{\includegraphics[width=\columnwidth]{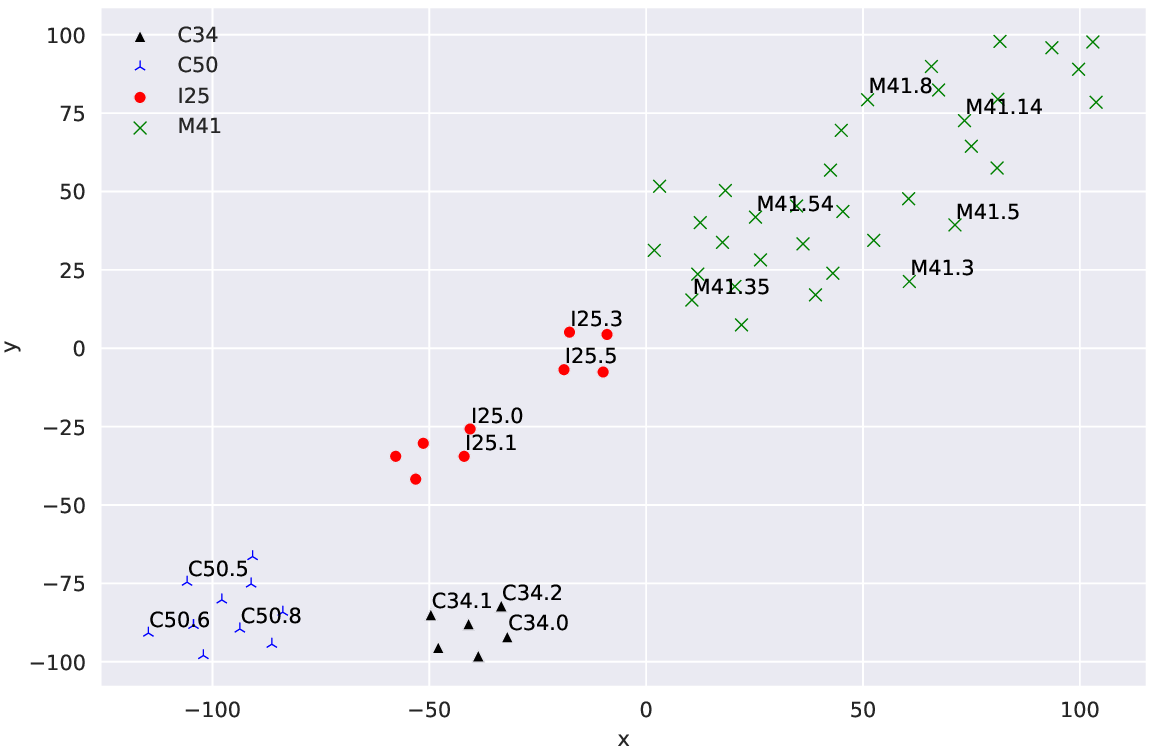}}
\caption{ICD-code embeddings projections from 4 disease groups: \emph{C34}, \emph{M41}, \emph{C50}, \emph{I25} (for readability purposes only some data points are labeled).} \label{figure_diseases}
\end{figure}

\begin{figure}[!t]
\centerline{\includegraphics[width=\columnwidth]{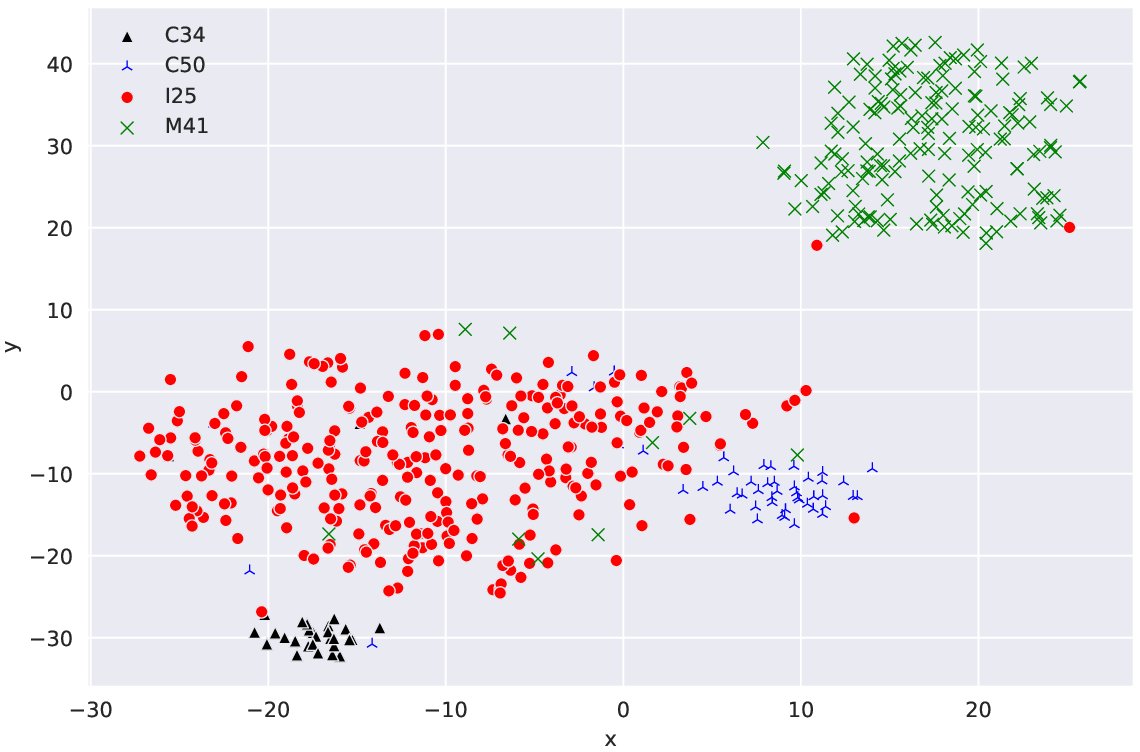}}
\caption{Sample of patient-embeddings with \emph{C34}, \emph{M41}, \emph{C50} or \emph{I25} ICD codes in their medical histories.} \label{figure_patient_diseases}
\end{figure}

For the life and health insurance industry it is crucial to differentiate diseases by disability or mortality risk. For example, the group of \emph{M41}-codes is definitely less risky compared to diseases in \emph{C34}, \emph{C50}, and \emph{I25} code groups. We try to show this difference in our embeddings with the t-SNE~\cite{Maaten2008} dimensionality reduction technique. The vectors of codes from the 4 above-mentioned groups projected to a plane are plotted on the Fig.~\ref{figure_diseases}. More importantly, this separation persists on the level of whole patient vectors. The Fig.~\ref{figure_patient_diseases} plots the excerpt of 578 random patient vectors, where each patient history contains code from one of the 4 above designated groups.

\begin{figure}[!t]
\centerline{\includegraphics[width=\columnwidth]{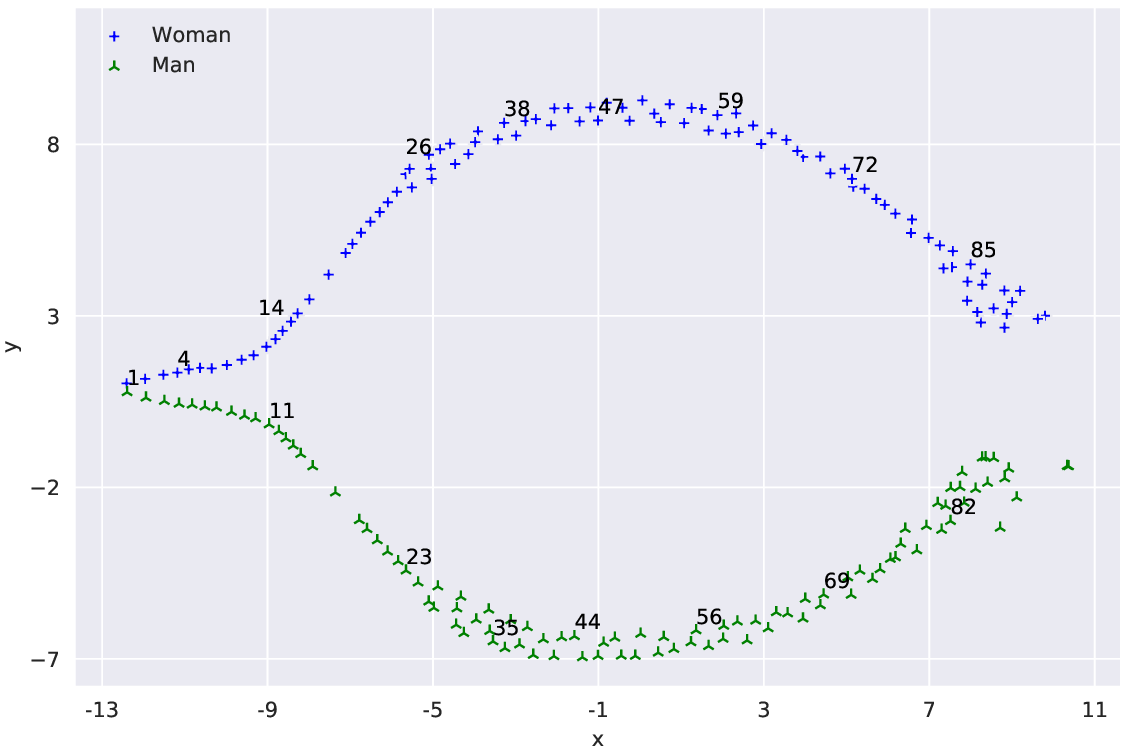}}
\caption{Generalized patient embeddings from the profile model for each gender and age.} \label{figure_patient_emb}
\end{figure}

Finally, we try to look at generalized patient representations, grouping patients by gender and age and presenting each group with an average of their embeddings. The result for men and women can be seen in Fig.~\ref{figure_patient_emb}. Each point on the plot is an average representation of thousands of patients. Again for readability, we labeled with age value only several data points. It is interesting to see clear gender separation and some age tendencies.

Here it is worth noting that the axis on Fig.~\ref{figure_diseases}, \ref{figure_patient_diseases} and \ref{figure_patient_emb} do not have any physical meaning. We cannot draw precise conclusions just from these plots as the t-SNE algorithm only tries to approximate relative object positions on the plain as close as possible to their positions in the original high-dimensional space. By these plots, we intended to provide only a general overview of the resulting embedding space and hope that it can adequately represent the medical concepts and patients.

\section{Profile Model Applications} \label{applications_sect}
As already mentioned, the profile model finds interesting scientific and practical applications in healthcare and related areas.

\subsection{Medical risk scoring} \label{risk_scoring}
Based on the developed model, we can infer the risk probabilities for a specific family of diseases (target) or even ICD-10 chapters:
\begin{equation}
\begin{aligned}
p(target \mid gender, age, hist)=\\\sum_{i}^{} p(icd_i\mid gender, age, hist), \label{probability_eq}
\end{aligned}
\end{equation}
where $icd_{i}$ is the i-th specific ICD code in the \emph{target} disease group.

\begin{figure}[!t]
\centerline{\includegraphics[width=\columnwidth]{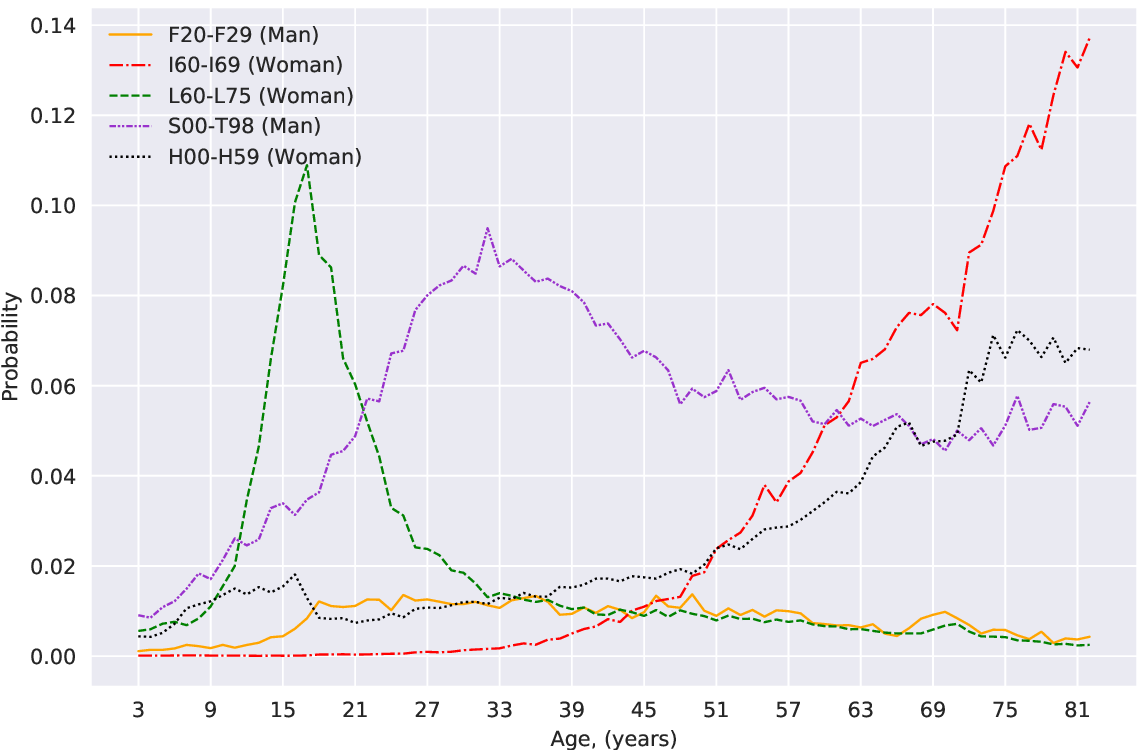}}
\caption{Risk curves for: \emph{F20-F29} (Schizophrenia, schizotypal and delusional disorders); \emph{I60-I69} (Cerebrovascular diseases); \emph{L60-L75} (Disorders of skin appendages); \emph{S00-T98} (Injury, poisoning and certain other consequences of external causes); \emph{H00-H59} (Diseases of the eye and adnexa).} \label{figure_risk}
\end{figure}

Namely, we can make up a set of patients varying only their mandatory parameters (leaving medical history empty), submit them to the model, and assess their target probabilities~\eqref{probability_eq}. Figure~\ref{figure_risk} shows age-dependent risk curves for five targets (disease groups). Again from this plot, one can note reasonable dependencies learned from data, like increased risk of skin disorders due to hormonal changes in a teenage body or peaking of cerebrovascular disease risks after 50 years of age, one of the leading causes of death in Europe~\cite{rapsomaniki2014blood,wilkins2017european}.

We can develop the idea with risk assessment even further, but first, let us verify some more properties of our model.

During the long history of medical practice and research, many findings about disease dependencies and their interconnection have been revealed. For example, the first column of Table~\ref{table_med_facts} lists three such known findings. More formally, we can rephrase these statements in terms of a trigger and target diseases, as shown in the second and third columns. What we want to check is the quantitative justification of those facts by our model.

\begin{table*}
\centering
\caption{Examples of medical findings and their quantitative assessment by the profile model.} \label{table_med_facts}
\begin{tabular}{m{5cm} >{\centering\arraybackslash}m{1cm} >{\centering\arraybackslash}m{2cm} >{\centering\arraybackslash}m{1.5cm} >{\centering\arraybackslash}m{.7cm} >{\centering\arraybackslash}m{1.7cm} >{\centering\arraybackslash}m{1.7cm} >{\centering\arraybackslash}m{.7cm}}
\hline
\bfseries Finding & \bfseries \begin{tabular}{@{}c@{}}Trigger\\disease\end{tabular} & \bfseries \begin{tabular}{@{}c@{}}Target\\diseases\end{tabular} & \bfseries \begin{tabular}{@{}c@{}}Gender,\\{[Age]}\end{tabular} & \bfseries \begin{tabular}{@{}c@{}}Group\\size, g\end{tabular} & \boldmath $p(target \mid trigger)$ & \boldmath $p(target \mid \neg trigger)$ & \bfseries Odds \\
\hline
Hypertension significantly increases the risk of more severe heart diseases~\cite{rapsomaniki2014blood} & I10 & I20 I21 I50 I64 & \begin{tabular}{@{}c@{}}Man, \\{[50; 60]}\end{tabular} & 152 & 0.00164538740 & 0.00040872230 & 4.03 \\
\hline
Gastritis can develop dangerous complications such as duodenal ulcer or stomach cancer~\cite{delaney1970gastric,kato1992atrophic} & K29 & K25 K26 C16 & \begin{tabular}{@{}c@{}}Women,\\{[50; 60]}\end{tabular} & 412 & 0.00020268999 & 0.00003274679 & 6.19 \\
\hline
Obesity leads to the diabetes, heart and breathing problems~\cite{bray2004medical} & E66 & E11 I10 G47 & \begin{tabular}{@{}c@{}}Man\&Women,\\{[30; 40]}\end{tabular} & 323 & 0.00047231422 & 0.00014405760 & 3.28 \\
\hline
\end{tabular}
\end{table*}

For each statement, let us fix the demographic parameters (column \emph{Gender, {[Age]}}) and select two groups of patients: the one with the trigger disease code in their medical histories and the other without such codes. In each of the three cases, we select the \emph{Group size} based on the available number of patients with the \emph{Trigger disease} in the validation set (see Section~\ref{Data}) and take the same number of random "healthy" patients for the opposite control group.

Next, given the \eqref{probability_eq} we estimate the mean probabilities of \emph{Target diseases} (risk values) for both groups (columns $p(target \mid trigger)$ and $p(target \mid \neg trigger)$ in the Table~\ref{table_med_facts}). As the final step we compute the \emph{Odds} ratio~\cite{Manning} as:
\begin{equation}
Odds = \frac{p(target \mid trigger)}{p(target \mid \neg trigger)}. \label{odds_eq}
\end{equation}

In our examples, all odds values are much greater than one, that is the probabilities of target diseases much greater in a group of patients with trigger diseases than in control group. This confirms initial statements and allows to conclude that profile model correctly assess known medical knowledge and facts.

\begin{table}
\centering
\caption{Medical findings discovered by the H2D method.} \label{table_med_hypothesis}
\begin{tabular}{p{2.2cm} l}
\hline
\bfseries \begin{tabular}{@{}l@{}}Parameters (Target,\\Gender, Age, g=152)\end{tabular} & \bfseries \begin{tabular}{@{}l@{}}Harbinger disease candidates\end{tabular} \\
\hline
\begin{tabular}{@{}l@{}}Malignant neoplasm\\of breast\\C50,\\Women,\\{[50; 60]} \end{tabular} & \begin{tabular}{@{}l@{}}G70.0 G70.2 G35 G62.1 G62.2 -- F06.28 F06.2\\F06.21 F06.08 F22.9 F00.14 F06.29 F33.11\\F70.19 F07.08 F20 F84.02 F06.01 F33.10 F01.9\\F01.3 F06.0 F20.0 F06.3 -- M32.1 M32.8 M34.0\\M86.6 M86.3 M34.9 M86.1 M33.2 M34.8 M86.4\\M06.0 M80.8 -- J90 J94.8 J94.9 J38.0 J93.0 --\\I05.2 I97.2 I89.0 I74.2 I73.0 I74.3 -- K74.6 K74.3\\K74.0 -- R18 R09.1 -- N87.2 N30.4 -- E06.2 --\\L97 \end{tabular} \\
\hline
\begin{tabular}{@{}l@{}}Malignant neoplasm\\of bronchus and\\lung\\C34,\\Man,\\{[50; 60]} \end{tabular} & \begin{tabular}{@{}l@{}}J90 J84.9 J94.9 J84.8 J86.9 J94.8 J85.2 J84.1\\J94.0 J86.0 J98.4 J47.0 J47 J85.1 J38.0 J44.9\\J93.0 J44.8 J15.3 J93.9 J43.9 J44.1 J44.0 J98.1\\J93.1 J93.8 J39.3 J15.9 J43.2 J45.9 J46 J16.8\\J13 J18.9 J81 J15.8 J30.3 -- T75.2 T34.5 -- A16.0\\A16.1 A15.0 A15.3 A16.2 A41.9 A15.9\\A15.2 -- F84.11 F20.0 F84.01 F07.01 F06.29\\F20 F07.82 F84.8 F06.20 F10.31 F22.9 F84.02\\F07.88 F06.2 F01.00  -- G40.1 G40.3 G40.0 G40.2\\G70.0 G41.9 G40.5 G70.2 G40.8 G35 G40.6\\G62.1 G40 -- M45.0 M86.3 M06.0 -- B22.7\\B20.7 -- I42.0 I26.9 I65.2 I50.0 I27.9 I71.4\\I26.0 I73.1 I70.0 I08.3 I73.0 I89.9 I71.6 -- Q33.0\\Q33.5 -- H35.5 H18.1 -- N18.5 N18.4 N18.0
\end{tabular} \\
\hline
\begin{tabular}{@{}l@{}}Melanoma and\\malignant neoplasms\\of skin\\C43-C44,\\Man\&Women,\\{[30; 60]} \end{tabular} & \begin{tabular}{@{}l@{}}G35 G40.2 G70.2 G70.0 -- L93.0 L94.0 L82 L10.0\\L57.0 -- E27.9 E22.0 E27.8 E23.6 E89.0 E22.1 -- \\
F21.3 F21 F31.7 F31.31 F31.6\\F01.3 F33.2 F84.11 F20.3 F33.11 F07.02 F84.5\\ F06.21 F01.80 F31.0 F20.8 F23.31 -- M32.1\\M45.0 M05.8 M06.8 M35.3 M24.4 M16.2\\M08.3 -- K50.0 K50.1 -- H02.7 H44.2 -- I05.2\\I78.1 -- B97.7
\end{tabular}\\
\hline
\begin{tabular}{@{}l@{}}Colorectal cancer\\C18-C20,\\Man,\\{[50; 60]} \end{tabular} & \begin{tabular}{@{}l@{}}K63.2 K62.1 K56.6 K56.5 K57.2 K50.1 K62.7\\K63.5 K66.0 K65.8 K51.3 K63.1 K50.0 K91.4 -- \\
Z93.3 -- N13.3 N13.1 -- G35 -- H35.5 -- T83.1
\end{tabular}\\
\hline
\begin{tabular}{@{}l@{}}Leukemia of\\unspecified cell type\\C95,\\Man\&Women,\\{[1; 10]} \end{tabular} & \begin{tabular}{@{}l@{}}Q21.3 -- F71.08 F72.02 F71.04 F72.08 F71.02 -- \\
G80.9 G80.4 G80.2 G80.0 G70.0 G80.3 G80.1\\G37.8 G71.0 -- N17.9 N17.8 N95.1 -- A41.9 -- \\
H47.2 -- I11.0 I20.8 -- K74.6 -- E25.0
\end{tabular}\\
\hline
\end{tabular}
\end{table}

That means that we can reverse the procedure of risk assessment and use the profile model to discover triggers or associate diseases for any target one. We call this procedure the \emph{H}arbinger \emph{D}isease \emph{D}iscovery method or \emph{H2D method} for short. The formal steps of such a method are listed in the Algorithm~\ref{hypothesis_gen_alg}.

\begin{algorithm}
\caption{H2D method implementation} \label{hypothesis_gen_alg}
    \hspace*{\algorithmicindent} \textbf{Input} $target, gender, age, g$ \Comment{\emph{target} disease for study, demographic parameters and \emph{Group size}}
    \begin{algorithmic}[1]
    
    \State $G \gets \Call{RandomG}{\neg target, gender, age, g}$
    \State $p_{control} \gets \Call{EstimateProbability}{G}$
    \ForAll{$icd_i$ in $V$} \Comment{V is the vocabulary of "tokens"}
        \State $G' \gets \Call{ModifyHist}{icd_i, G}$
        \State $p' \gets \Call{EstimateProbability}{G'}$
        \State $odds \gets \frac{p'}{p_{control}}$ \Comment{\eqref{odds_eq}}
    \EndFor
    
    \State   
    \Function{EstimateProbability}{$G$}
        \ForAll{$patient_j$ in $G$}
            \State $p_j(target) \gets p(target \mid {gender}_j, {age}_j, {hist}_j)$ \Comment{\eqref{probability_eq}}
        \EndFor
        \State \Return $\frac{1}{g}\sum_{}^{} p_j$
    \EndFunction
    \end{algorithmic}
\end{algorithm}

First, we select a random group \emph{G} of "healthy" patients according to the input parameters and estimate the target probability value ($p_{control}$) for this group. We are interested in the effect of a tested ICD ($icd_i$) on the target probability. That is done by adding an $icd_i$ to all patients' histories thus creating a modified group ($G'$). What is left is to reassess the target probability ($p'$) on such a simulated group and compute the \emph{odds} ratio~\eqref{odds_eq} for the tested ICD code. In the end, we get odds values for all (6,986) ICD codes available to the profile model. By selecting codes only with values greater than 2 (i.e. that at least doubles the target probability), we obtain the most probable harbingers or strongly associated diseases for the target one. Table~\ref{table_med_hypothesis} presents the result of the H2D method for five common cancer diseases. It is worth noting that we intentionally exclude from the resulting lists the whole ICD-10 chapter on neoplasms itself because its relationship with the designated targets is too obvious.

Looking through the results we stumbled on some non-obvious patterns. For example, to our surprise, there are a lot of mental and psychological disorders (the \emph{F*} group of codes) associated with the target \emph{C50}. To clear our doubts about this finding we resorted to the analysis of related research literature. And even to more of our surprise found several studies exactly on the subject. In \cite{lu2020shared} the epidemiological bidirectional link between breast cancer and schizophrenia has been shown. Similarly, the authors of~\cite{zhuo2018association} demonstrate that in comparison with the general female population, women with schizophrenia are at a higher risk for the incidence of breast cancer. At the same time, some of the proposed hypotheses (e.g. the association between myasthenia (\emph{G70*}) and cancer) already have been verified and rejected~\cite{pedersen2014myasthenia}.

The above allows us to state that our proposed method generates curious hypotheses about disease interactions to which scientific interest has been confirmed by experimental research. Further analysis of Table~\ref{table_med_hypothesis} lies beyond the scope of the current paper, and we will leave it to the interested parties. We hope that our H2D method will be inspiring for medical researchers and allow them to find novel or underexplored research directions.

\subsection{Insurance scoring} \label{ins_scoring}
Finally, we apply the above-discussed model in one of our current projects. The stakeholders from an insurance company want to rebuild and automate part of their scoring pipeline. Insurance risk can vary widely for different customers depending on many factors. An accurate predictive model for risk assessment leads to an optimal personalized charge which is beneficial for a company and customer. The core modeling object in such a problem is an \emph{Application for insurance Policy, AP}. Formally it requires building a model \emph{f} which for a given \emph{AP} predicts the risk value \emph{r}: $f(AP)=r$. The risk \emph{r} is defined as the following: claim of insurance sum during the first year period, meaning it is a binary event - whether an \emph{AP} resulted in a loss. Therefore it was decided to address the problem as a binary classification task.

At our disposal, we had the dataset of historical \emph{AP}s for several years (see Section~\ref{Data}). From these data it was concluded that each \emph{AP} object consists of two major parts: \emph{Applicant, A} and \emph{Policy, P}. The \emph{A}-part contains person-related features such as gender, age, diseases anamnesis in the form of free text and ICD-10 codes. The \emph{P}-part includes all contract-related features (insurance period and product type, insured sum and currency, region, etc.). Also according to historical \emph{DIns} data the binary target is very skewed with less than 1\% of claims.

Given \emph{AP} features separation we can use patient embeddings to fully represent the whole \emph{A}-part in a unified fashion. Accounting on the properties of our embedding space we can naturally process applicants with rare or even unseen disease anamnesis. For instance, there are many applicants with the \emph{I25} disease and mostly they are in the high-risk group. Suppose in a new \emph{AP} we encountered with the \emph{I21} diagnosis code unseen in \emph{DIns} data. But both \emph{I25} and \emph{I21} relate to heart disease and their vectors are close together in the embedding space (as we learned from medical data) so we can more precisely assess the risk in this case.

It is worth noting that due to the early stage of this project development, substantial part of diagnosis-related features is still in the process of consolidation. In the case of an empty anamnesis feature field, we simply fall back to generalized patient representations.

Also, the important requirement to the model was the stability of predictions in time and interpretability. For these reasons, we choose the Logistic Regression (LR) model~\cite{bishop2006pattern} as above mentioned \emph{f} function. This model is compared under two schemes: \emph{Base} and \emph{Replacement}. In the \emph{Base} scheme only insurance data were used. \emph{A}-part features were one hot encoded before concatenation to \emph{P}-features. In the \emph{Replacement} scheme the \emph{A}-part features group was replaced by 256-dimensional patient embeddings.

The models' performance was measured by the ROC AUC metric~\cite{Manning}. Table~\ref{table_auc} shows by-month and average metric values under both mentioned schemes for the validation year. It can be seen that both models are stable by months. The \emph{Replacement} scheme yields solid metric improvements ranging from 0.5\% to 3.4\% and 1.82\% in the year average. To check statistical significance of the improvement, we apply Student's t-test. Given the data from Table~\ref{table_auc} the result statistic value is 2.77, which is more than critical value 2.07 (at the significance level of 0.05) thus differences in ROC AUC averages are statistically significant. This allows us to conclude that using patient embeddings helps in the given task. We hope to further improve the metric as the full insurance features become available.

\begin{table}
\centering
\caption{Validation ROC AUC metrics of the scoring model, (\%).} \label{table_auc}
\begin{tabular}{rcc}
\hline
\multirow{2}{*}{\bfseries Month}
    &   \multicolumn{2}{c}{\bfseries Scheme}\\
    \cline{2-3}
    & \bfseries Base & \bfseries Replacement \\
\hline
January & 71.6 & 75.0\\
February & 72.4 & 74.1\\
March & 71.9 & 73.6\\
April & 70.4 & 72.6\\
May & 67.1 & 69.1\\
June & 71.7 & 72.2\\
July & 70.4 & 71.8\\
August & 72.5 & 74.0\\
September & 71.2 & 72.8\\
October & 69.5 & 71.6\\
November & 71.8 & 73.6\\
December & 69.4 & 71.6\\
\hline
Average & 70.83 & \bf 72.65\\
\hline
\end{tabular}
\end{table}

\subsubsection{A note on the model deployment}
To integrate the developed model in the current scoring pipeline, we prototyped RESTful web service using Flask framework~\cite{grinberg2018flask}. The service receives applicant data from insurance software, runs the transformer neural network to get an applicant embedding, applies the final LR model $f(AP)$, and returns the result. As we plan to assess this model performance through A/B testing the service also logs each query in the Postgres database for further analytical purposes.

When designing a service for a real-time use case scenario, the response time is a primary consideration. In the above-listed steps, the most time-consuming one is the transformer-model inference. It depends on the computation device specification, for instance with NVIDIA Tesla V100 GPU it takes on average $6.6~\si{\milli\second}$ to process a single query. The CPU version is roughly ten times slower - $68.6~\si{\milli\second}$ per request, but for the pilot period, even this performance is enough to manage our current workload. For smooth deployment, we packed the whole service into a docker image, separating it from the database. In such a way, the application allows horizontal scaling under increasing load by running more processing containers that share a single database located in another container.

Another concern to address is the stability monitoring of the hosted model. For this, we selected Population Stability Index (PSI)~\cite{siddiqi2012credit}, which measures the difference between model score distributions on development and production sets. Weekly monitoring of the PSI allows for quickly detecting an unexpected model behavior to reduce financial and time loss.

\section{Conclusions} \label{Conclusions}
We presented the profile model for working with medical data on the level of holistic patient histories through embedding space built with the neural network transformer architecture. Model analysis revealed that it automatically learns several plausible medical patterns and adequately preserves relation between concepts. Our experiments on the MIMIC-III data showed that the inclusion of gender and age-specific information without positional embeddings allows us to achieve the new state-of-the-art result in the next diagnosis prediction task. We release the code of that experiment, hoping it will be useful for other EHR-related research. Next, we will plan to extend the described representation learning approach by trying to incorporate more medical concepts into the model.

Besides, we showed two applications of the developed model. The first one relates to the medical research methodology. The proposed H2D method allows the revealing of disease-associated hypotheses. For example, we present strong-associated hypotheses set for five types of cancers. We hope such a data-driven method will lead the medical community to more focused and meaningful research.

The second application relates to the knowledge transferring between domains. We showed how patient embeddings representation could be extracted from the profile model and successfully applied to the practical task from the insurance domain. On the validation data, we obtained a stable ROC AUC metric improvement of 1.82\%. To verify the benefits of this model, we plan to pilot it in the production environment. From our point of view, the deployment of such a model has a couple of major challenges for both of which we propose reasonable solutions.

% \bibliographystyle{splncs04}
% \bibliography{bibliography}

% \begin{thebibliography}{00}
% \end{thebibliography}

\end{document}